\documentclass[twocolumn,letter]{jpsj3}

\usepackage{txfonts}
\usepackage{bm}
\usepackage{color}%

\def\Vec#1{\bm{#1}}

\newcommand{\imu}{{\rm i}}

\begin{document}


\title{Quasiclassical Treatment and Odd-parity/Triplet Correspondence in Topological Superconductors}

\author{Yuki \surname{Nagai}, Hiroki \surname{Nakamura}, and Masahiko \surname{Machida}}
\inst{
\address{CCSE, Japan  Atomic Energy Agency, 5-1-5 Kashiwanoha, Kashiwa, Chiba 277-8587, Japan} 
}

\date{\today}
             
\abst{

We construct a quasiclassical framework for topological superconductors with the strong spin-orbit coupling such as Cu$_{x}$Bi$_{2}$Se$_{3}$. 
In the manner of the quasiclassical treatment, 
decomposing the slowly varying component from the total quasi-particle wave function, 
the original massive Dirac Bogoliubov-de Gennes (BdG) Hamiltonian derived from the tight-binding model represented by $8 \times 8$ matrix is reduced to $4 \times 4$ one. 
The resultant equations are 
equivalent to Andreev-type equations of singlet or triplet superconductors, in which the apparent spin-orbit coupling vanishes. 
Using this formalism, we find a fact that the odd-parity superconductivity in topological superconductors turns to the spin-triplet 
one. 
Moreover, in terms of the quasiclassical treatment, we show that the topologically-protected zero-energy states in topological superconductors has the correspondence 
to the Andreev bound states established in a long history of studies for the unconventional superconductors. 
This clearly indicates that low-energy non-trivial superconducting properties in the topological superconductors 
can be analyzed by the established theoretical descriptions on the spin-triplet superconductors. 
}
%
\kword{superconductivity, topological superconductors, spin-orbit coupling, quasiclassical theory}
\maketitle
Topological superconductors have attracted much attention, because of their interesting states protected by the non-trivial topological invariants. A large number of experimentalists have intensively explored clear evidence of the topological superconductivity by various tools, while theorists have challenged to make convenient theoretical frameworks to draw their non-trivial nature\cite{Bernevig15122006,PhysRevLett.105.266401,PhysRevB.76.045302,PhysRevLett.98.106803,RevModPhys.82.3045,PhysRevLett.95.146802,Konig02112007,PhysRevLett.105.14 6801,PhysRevB.75.121306,PhysRevB.81.041309,PhysRevLett.105.136802}.
One of the significant phenomena caused by non-trivial topology is that gapless states, including the zero-energy states, are robustly formed at a boundary between the phases with different topological invariants. In the case of topological superconductors, the zero-energy quasiparticles are locally bound also in the center of a quantized magnetic vortex. Such robust zero-energy bound states are expected to be applicable to quantum computing as the Majorana fermions\cite{Teo}.

In unconventional superconductors, the theoretical description on the bound states at their surfaces has been intensively investigated. Consequently, the emergence of the zero-energy states at surfaces has been regarded as clear evidence of the unconventional superconductivity\cite{KopninText,TanakaKashiwaya,Kashiwaya}. 
These bound states called the Andreev bound states form when the gap function changes its sign along the quasiparticle scattering path at the boundary. Therefore, the momentum-independent gap functions never generate the Andreev bound states at a boundary, since they cannot change its sign through the scattering process. 
However, the three-dimensional topological superconductor Cu$_{x}$Bi$_{2}$Se$_{3}$ has the zero-energy bound states at surfaces even with the momentum-independent and fully-gapped odd-parity gap function due to the on-site interaction\cite{PhysRevLett.107.217001,PhysRevB.86.064517}. 
According to the above discussion, this bound state would be different from the Andreev bound state.

On the other hand, we reported that the topological superconductor with momentum-independent gap function has a point-node-like excitation and the strong anisotropy of the thermal conductivity below $T_{\rm c}$\cite{NagaiThermal},  due to the strong momentum-dependent spin-orbit couplings. These results imply that the gap functions of topological superconductors have effective momentum dependence caused by the spin-orbit coupling. 
If we construct an effective theory with apparently momentum-dependent gap functions for topological superconductors, the techniques to analyze the Andreev bound states in unconventional superconductors can be applied to the surface bound states of topological superconductors.


In this paper, we propose an effective theory with the $2 \times 2$ spin-triplet and momentum-dependent order-parameter instead of the original theory for topological superconductors with the $4 \times 4$ odd-parity order-parameter,  in terms of the quasiclassical treatment of superconductivity. 
We argue a theoretical correspondence between 
the topologically protected bound state and the zero-energy Andreev bound state. 
The present quasiclassical treatment can
 map the topological superconductivity 
into the spin-triplet one eliminating the spin-orbit coupling. 
According to this treatment, the protected surface states are reinterpreted as the 
the Andreev zero-energy bound states established in a long history of studies for
the unconventional superconductors.  
Thus, 
we reveal that the topological superconducting nature brought about by the spin-orbit coupling can be explained by spin-triplet superconductivity without the coupling. 

The quasiclassical theory is successful in the weak-coupling Bardeen-Cooper-Schrieffer (BCS) type of superconductivity \cite{KopninText}. 
The theoretical framework is grounded on the fact that the coherence length $\xi$ is sufficiently longer than 
the Fermi wave length $1/k_{\rm F}$, i.e., $ \xi k_{\rm F} \gg 1$ which expresses a typical scale difference in 
the weak-coupling superconductivity. 
Thus, the quasi-particle wave function can be approximated as a product of the fast ($1/k_{\rm F}$) and slowly ($\xi$) spatially-variational functions, and 
the quasiclassical theory concentrating only on the slow component offers the significantly reduced models compared to  
the original Bogoliubov de Gennes (BdG) equations. Mathematically, the second-order differential eigen  
equations are reduced to the {\it first}-order differential equations on 
the {\it one}-dimensional line called the trajectory\cite{KopninText,Volovik,NagaiJPSJ:2006,Eilenberger}.  
So far, various types of the analytical and numerical techniques on the quasiclassical theory
have been developed and successfully applied to a tremendous number of conventional and unconventional superconductors\cite{KopninText,Volovik,NagaiJPSJ:2006,Eilenberger,NagaiMeso,Miranovic,Melnikov,NagaiPRL,NagaiCe,Graser,Iniotakis}. 

The quasiclassical treatment of the topological superconductors itself is very important in order to treat the inhomogeneous systems. 
Since the BdG framework requires huge-scale numerical computations for a diagonalization the BdG Hamiltonian, the quasiclassical approximation has been utilized with significant 
computational cost reduction which enables to access several inhomogeneous situations like a vortex, its lattice, an interface with non-superconducting blocks, and so on. 
Indeed, the case of the topological superconductor with the strong spin-orbit coupling such as Cu$_{x}$Bi$_{2}$Se$_{3}$ is 
the most prominent example.
The minimum BdG model for topological superconductors 
demands two orbital degrees of freedom coupled with two spin degrees\cite{PhysRevLett.107.217001,PhysRevLett.105.097001,PhysRevB.83.134516,Yip,Hashimoto},  and the resultant 
BdG equations are represented by an 8$\times$8 Hamiltonian matrix, 
much larger than the conventional 2 $\times$ 2 one.

In this paper, we construct a reasonable quasiclassical framework on the topological superconductors. 
Starting with the massive Dirac BdG Hamiltonian describing   
the topological superconductors\cite{NagaiThermal,NagaiMajo},  
we present that the derived quasiclassical BdG equations become equivalent 
to the conventional linearized BdG (Andreev) equations for 
spin singlet or triplet gap function depending on the original gap types. 
Especially, we point out that all the possible 
odd-parity order-parameters can be mapped onto 
the spin-triplet ones through the present quasiclassical treatment. 
This fact clearly indicates that any low-energy topological superconducting properties in the odd-parity order-parameters 
are linked to those of the triplet superconductivity.


Now, let us begin 
to construct a quasiclassical formalism for topological superconductors. 
The typical topological superconductor Cu$_{x}$Bi$_{2}$Se$_{3}$ can be described by 
the massive Dirac-type BdG Hamiltonian, which includes spin-orbit coupling, expressed as\cite{NagaiThermal,NagaiMajo}  
%
\begin{equation}
H = \int d{\bm r}
\left(\begin{array}{cc}\bar{\psi}({\bm r}) & \bar{\psi}_{\rm c}({\bm r})\end{array}\right)
\left(\begin{array}{cc}\hat{H}^{-}({\bm r})  & \Delta^{-}({\bm r}) \\
\Delta^{+}({\bm r})& \hat{H}^{+}({\bm r}) \end{array}\right)
\left(\begin{array}{c}\psi({\bm r}) \\
\psi_{\rm c}({\bm r})
\end{array}\right), \label{eq:hami}
\end{equation}
where 
\begin{equation}
\hat{H}^{\pm}({\bm r}) = M_{0}  - \imu  \partial_{x} \gamma^{1} - \imu \partial_{y} \gamma^{2}
- \imu \partial_{z} \gamma^{3} \pm \mu \gamma^{0} .\label{eq:dirac}
\end{equation}
Here, $\gamma^{i}$ is a $4 \times 4$ Dirac gamma matrix,  which can be described as $\gamma^{0} = \hat{\sigma}_{z} \otimes 1$,
$\gamma^{i = 1,2,3} = i \hat{\sigma}_{y} \otimes \hat{s}_{i}$, and $\gamma^{5} = \hat{\sigma}_{x} \otimes 1$ 
with $2 \times 2$ Pauli matrices $\hat{\sigma}_{i}$ in the orbital space and $\hat{s}_{i}$ in the spin space, 
$\psi(\bm r)$ is the Dirac spinor, $\bar{\psi}(\Vec{r}) \equiv \psi^{\dagger}(\Vec{r})\gamma^{0}$, $\bar{\psi}_{c}(\Vec{r}) 
\equiv \psi_{c}^{\dagger} \gamma^{0}$, and $\psi_{c} \equiv {\cal C} \bar{\psi}^{T}$, 
where ${\cal C} (\equiv i \gamma^{2} \gamma^{0})$ is the representative matrix of charge conjugation. 
$\Delta^{-}$ is the gap function and $\Delta^{+} \equiv \gamma^{0} (\Delta^{-})^{\dagger} \gamma^{0}$.
Considering only the on-site pairing interaction, the possible gap forms are reduced into six types of functions 
as seen in Table I.\cite{NagaiThermal} 
From the Hamiltonian Eq.~(\ref{eq:hami}), the correspondent $8 \times 8$ BdG equations are given as
\begin{equation}
\left(\begin{array}{cc} \hat{h}_{0}({\bm r}) - \mu  & \hat{\Delta}({\bm r})\\
 \hat{\Delta}^{\dagger}({\bm r}) & \hat{h}_{0}({\bm r}) + \mu \end{array}\right)
\left(\begin{array}{c}
u({\bm r}) \\
u_{\rm c}({\bm r})
\end{array}\right)
= E \left(\begin{array}{c}
u({\bm r}) \\
u_{\rm c}({\bm r})
\end{array}\right), \label{eq:bdgdirac}
\end{equation}
where $\gamma^{0} \hat{H}^{\pm} = \hat{h}_{0} \pm \mu$, and $\hat{\Delta} = \gamma_{0} \Delta^{-}$.
Note that $v$ in the conventional Nambu eigen state form, $(u, v)^T$ is related to $u_{c}$ 
as $v \equiv i \gamma^{2} u_{c}$.  

The quasiclassical theory is founded on an assumption 
that the coherence length $\xi$ is much longer than the  
Fermi wave length $1/k_{\rm F}$ ($\xi k_{\rm F} \gg 1$)\cite{Volovik}.  
This assumption is valid, when the order parameter amplitude $|\Delta_{0}|$ 
is much smaller than the Fermi energy $E_{\rm F}$ ($|\Delta_{0}|/E_{\rm F} \ll 1$), and 
this condition is fully fulfilled in BCS weak-coupling superconductivity.
In this theory, the wave function is expressed by a 
product of the fast oscillating one characterized 
by the Fermi momentum $p_{\rm F}$ and the slowly varying one by the coherence length $\xi$.  
The quasiclassical solution of the BdG equations is given as 
\begin{equation}
\left(\begin{array}{c}
u({\bm r}) \\
u_{\rm c}({\bm r})
\end{array} 
\right) \sim 
\left(\begin{array}{c}
 \Vec{u}_{1}^{\rm N}(\Vec{r},\Vec{p}_{\rm F})  f_{1}(\Vec{r},\Vec{p}_{\rm F})+ \Vec{u}_{2}^{\rm N}(\Vec{r},\Vec{p}_{\rm F})  f_{2}(\Vec{r},\Vec{p}_{\rm F})  \\
\Vec{u}_{{\rm c} 1}^{\rm N}(\Vec{r},\Vec{p}_{\rm F})  g_{1}(\Vec{r},\Vec{p}_{\rm F}) +\Vec{u}_{{\rm c} 2}^{\rm N}(\Vec{r},\Vec{p}_{\rm F})  g_{2}(\Vec{r},\Vec{p}_{\rm F}) 
\end{array}\right), \label{eq:appro}
\end{equation}
where 
$f_{i}$ and $g_{i}$ correspond to slowly varying components, 
$\Vec{u}_{i}^{\rm N}$, $\Vec{u}_{{\rm c} i}^{\rm N}$ are 
the fast oscillating function and adopeted as 
normal-state eigenvectors satisfying the eigen-equations,  
\begin{eqnarray}
 \hat{h}_{0}(\Vec{r}) 
\Vec{u}_{i}^{\rm N}(\Vec{r},\Vec{p}_{\rm F}) &= \mu \Vec{u}^{\rm N}_{i}(\Vec{r},\Vec{p}_{\rm F}) , \\
\hat{h}_{0}(\Vec{r}) 
\Vec{u}_{{\rm c} i}^{\rm N}(\Vec{r},\Vec{p}_{\rm F}) &= -\mu \Vec{u}_{{\rm c} i}^{\rm N}(\Vec{r},\Vec{p}_{\rm F})  .
\end{eqnarray}
Here, the chemical potential is supposed to be larger than the mass $(\mu > M_{0})$.
The eigenvectors are given as 
\begin{equation}
\Vec{u}_{i}^{\rm N}(\Vec{r},\Vec{p}_{\rm F}) = 
c e^{\imu \Vec{p}_{\rm F} \cdot \Vec{r}} 
 \left(\begin{array}{c}
\chi_{i}  \\
\frac{\Vec{p}_{\rm F} \cdot \Vec{\sigma}}{E_{0} + M_{0}} \chi_{i} 
\end{array}\right), \label{eq:ui}
\end{equation}
and 
\begin{equation}
\Vec{u}_{{\rm c} i}^{\rm N}(\Vec{r},\Vec{p}_{\rm F}) = c e^{\imu \Vec{p}_{\rm F} \cdot \Vec{r}} 
 \left(\begin{array}{c}
-  \frac{\Vec{p}_{\rm F} \cdot \Vec{\sigma}}{E_{0} + M_{0}} \chi_{i}  \\
\chi_{i} 
\end{array}\right),
\end{equation}
where, $\chi_{1}^{\rm T} = (1,0)$, $\chi_{2}^{\rm T} = (0,1)$, and $c \equiv \sqrt{(E_{0}+M_{0})/2E_{0}}$ obtained 
by the normalized condition $u^{{\rm N} \dagger}_{i} u_{i}^{\rm N} $ = 1 and $E_{0} = \mu =  \sqrt{M_{0}^{2}+\Vec{p}_{\rm F}^{2}}$.
These solutions $\Vec{u}_{i}^{\rm N}$ and $\Vec{u}_{{\rm c} i}^{\rm N}$ are well known as the free particle and anti-particle solutions 
in the Dirac equation in high energy physics, respectively.  
Here, we note that $f_{1}$ and $f_{2}$ have dominant weights on 
the up and the down spins, respectively, in the non-relativistic limit $|\Vec{p}_{\rm F}|/M_{0} \ll 1$.  
With the use of the above wave functions, we reach $4 \times 4$ matrix eigenvalue 
problem with respect to four functions $(f_{1},f_{2},g_{1},g_{2})$ from $8 \times 8$ BdG equations.
The diagonal 
blocks in the BdG Hamiltonian (Eq.~(\ref{eq:bdgdirac})) are converted as 
\begin{equation}
f_{i}^{\ast} \Vec{u}_{i}^{\rm N}(\Vec{r},\Vec{p}_{\rm F})^{\dagger} (\hat{h}_{0} - \mu) \Vec{u}_{j}^{\rm N}(\Vec{r},\Vec{p}_{\rm F}) f_{j}
= 
 -f^{\ast}_{i}(\Vec{r}) \imu \Vec{v}_{\rm F} \cdot \Vec{\nabla} f_{j}(\Vec{r}) \delta_{ij}, 
\end{equation}
where we use the relation $\Vec{v}_{\rm F} \equiv \partial E_{0} /\partial \Vec{p}_{\rm F} = \Vec{p}_{\rm F}/E_{0}$. 
The diagonal term includes $\Vec{v}_{\rm F} \cdot \Vec{\nabla}$, which 
is well known as the differential operator in the conventional quasiclassical theory\cite{Volovik,Eilenberger, NagaiJPSJ:2006}. 
We note that the diagonal term does not have the spin-orbit coupling. 
The conversions of the off-diagonal blocks depend 
on the type of the gap functions described below. 
Eventually, we have effective $4 \times 4$ quasiclassical BdG  equations represented as 
\begin{equation}
\left(\begin{array}{cc} - \imu \Vec{v}_{\rm F} \cdot \Vec{\nabla}  & \hat{\Delta}_{\rm eff}({\bm r},\Vec{p}_{\rm F})\\
\hat{\Delta}_{\rm eff}^{\dagger}({\bm r},\Vec{p}_{\rm F}) &  \imu \Vec{v}_{\rm F} \cdot \Vec{\nabla} \end{array}\right)
\left(\begin{array}{c}
\Vec{f}({\bm r},\Vec{p}_{\rm F}) \\
\Vec{g}({\bm r},\Vec{p}_{\rm F})
\end{array}\right)
= E \left(\begin{array}{c}
\Vec{f}({\bm r},\Vec{p}_{\rm F}) \\
 \Vec{g}({\bm r},\Vec{p}_{\rm F})
\end{array}\right), \label{eq:quasi}
\end{equation}
where $\Vec{f}^{T} = (f_{1},f_{2})$, and $\Vec{g}^{T} = (g_{1},g_{2})(- i \sigma_{y})^{T}$.
The gap functions are converted into 
$\hat{\Delta}_{\rm eff} \equiv \hat{\Delta}_{\rm quasi} i \sigma_{y}$, where $(\hat{\Delta}_{\rm quasi})_{ij} 
\equiv \Vec{u}_{i}^{\rm N}(\Vec{r},\Vec{p}_{\rm F})^{\dagger} \hat{\Delta}(\Vec{r}) \Vec{u}_{{\rm c}j}^{\rm N}(\Vec{r},\Vec{p}_{\rm F})$. 
We multiply $i \sigma_{y}$ by $\Vec{g}$ to obtain the conventional BdG form. 
All the converted gap functions are listed in Table \ref{table:1}. 
All odd-parity gap functions are converted to spin-triplet ones, while all even-parity become spin-singlet ones. 
\begin{table*}[t]
\caption{
The correspondence between the original BdG gap functions $\hat{\Delta}^{-}$ and the effective ones 
$\hat{\Delta}_{\rm eff}(\Vec{p}_{\rm F})$
in quasiclassical theory.
``P-scalar'' denotes a pseudo scalar whose parity is 
odd and ``$i$-polar'' denotes a polar vector pointing the $i$ direction in four dimensional space.}
\label{table:1}
\begin{center}
\begin{tabular}{lcccccl}
\hline
&$\hat{\Delta}^{-}$  &Parity &$\hat{\Delta}_{\rm eff}(\Vec{p}_{\rm F})$ &Correspondent description  \\
\hline
Scalar & $\gamma^{5}$ &$+$ &$ \imu \sigma_{y}$ &singlet\\
$t$-polar & $\gamma^{0} \gamma^{5}$ &$+$ & $M_{0}  \imu \sigma_{y}/E_{0}$ &singlet \\
P-scalar & $1$ & $-$ & $\imu (\Vec{p}_{\rm F} \cdot \Vec{\sigma} ) \sigma_{y}/E_{0}$ &triplet: $\Vec{d} = (v_{x},v_{y},v_{z})$\\
$x$-polar & $\gamma^{1}\gamma^{5}$& $-$ & $\imu (\Vec{p}_{\rm F} \times \Vec{\sigma})_{x}  \sigma_{y}/E_{0} $
&triplet: $\Vec{d} = (0,-v_{z},v_{y})$
\\
$y$-polar  & $\gamma^{2}\gamma^{5}$ & $-$ & $\imu (\Vec{p}_{\rm F} \times \Vec{\sigma})_{y}  \sigma_{y}/E_{0} $
&triplet: $\Vec{d} = (v_{z},0,-v_{x})$
\\
$z$-polar & $\gamma^{3}\gamma^{5}$ & $-$& $\imu (\Vec{p}_{\rm F} \times \Vec{\sigma})_{z}   \sigma_{y}/E_{0} $ 
&triplet: $\Vec{d} = (-v_{y},v_{x},0)$\\
\hline
\end{tabular}
\end{center}
\end{table*}
As an example exhibited in Table \ref{table:1}, the pseudo-scalar order parameter is equivalent to 
the spin-triplet order parameter whose $\Vec{d}$-vector rotates in momentum space ($\Vec{d}(\Vec{k}) \propto \Vec{p}$).\cite{Sigrist} 
The polar vector types also correspond to spin-triplet order parameters characterized by the $\Vec{d}$-vectors\cite{ref} 
shown in Table \ref{table:1}. 
It should be noted that a pseudo-scalar type gap function is equivalent to that of B-phase in superfluid $^{3}$He.\cite{Leggett,Murakawa} 
Thus, one can observe the various kinds of interesting phenomena predicted in terms of studies of the B-phase in topological superconductors with a strong spin-orbit coupling. 
We address that the Eilenberger, Usadel, and Ginzburg-Landau equations\cite{KopninText} are also derived by our present treatment. 
The details will be shown elsewhere.


Here, we comment on the momentum dependence of 
the mass term $M_{0} \rightarrow M(\Vec{p}) = M_{0} + M_{1} \Vec{p}^{2}$ usually 
used in the topological insulators. 
The sign of $M_{0}/M_{1}$ determines whether the system is the topological insulator or not\cite{Yamakage}. 
Then, one can also obtain the quasiclassical BdG equations, by replacing the velocity $\Vec{v}_{\rm F}$ and $E_{0}$ by 
$\Vec{v}_{\rm F}'( \equiv \partial E/\partial \Vec{p}_{\rm F}) = (1 + 2 M(\Vec{p}_{\rm F}) M_{1}) \Vec{p}_{\rm F}/E$ 
and $E = \sqrt{M^{2}(\Vec{p}_{\rm F}) 
+ \Vec{p}^{2}_{\rm F}}$, respectively. 
We note that there are two Fermi momenta $\Vec{p}_{\rm F,1}$ and $\Vec{p}_{\rm F,2}$ in the same momentum direction when the Fermi velocity $\Vec{v}_{F 1}'(\Vec{p}_{\rm F,1}) < 0$. 
The bound states can form at each Fermi momenta in this ``two-band'' superconductor. 
Thus, in the present quasiclassical theory, 
The energy spectrum of bound states depends on the material parameters $M(\Vec{p}_{\rm F})$ and $M_{1}$, 
as discussed by Yamakage {\it et al}.\cite{Yamakage}.

Let us demonstrate the correspondence between the topologically protected bound states and the zero-energy 
Andreev bound states. 
The Andreev bound states occur when the sign of the gap function changes through the scattering process. 
Since the spin-triplet case has been extensively investigated about the Andreev bound state, one can easily draw 
new features of the topological superconductors through the previous 
rich knowledge due to the correspondence.  
The first demonstration is the topologically-protected bound states at an interface, 
and the second is those in a vortex. 
It is well known that 
a topological superconductor with the fully-gapped gap function called the pseudo-scalar type 
has the topologically-protected surface gapless states\cite{NagaiThermal}. 
In terms of the Andreev bound states, the zero-energy bound states form when the sign of the gap function 
changes through the scattering process\cite{TanakaKashiwaya, Kashiwaya,PhysRevB.23.5788}. 
The correspondent effective $\Vec{d}$-vector for the pseudo-scalar type gap function is 
proportional to the momentum $\Vec{d}(\Vec{p}) \propto \Vec{p}$ as shown in Table \ref{table:1}. 
Thus, the sign change of the gap function always occurs in the backward scattering ($\Vec{p}_{\rm out} = - \Vec{p}_{\rm in}$), where the momentum $\Vec{p}_{\rm in(out)}$ is that of the initial (final) quasiparticle states. 
We note that the sign of the $\Vec{d}$-vector changes through the backward scattering $\Vec{d}(\Vec{p}_{\rm in}) = - \Vec{d}(\Vec{p}_{\rm out})$ in all topological odd-parity gap functions as shown in Table \ref{table:1}. 
In the case of the polar-vector type superconductors, it should be noted that 
there is no gapless state in a specific direction as shown in Refs.\citen{PhysRevLett.107.217001,Hashimoto, Yip}. 
This is naturally explained by the fact that the sign of the gap function does not change if the surface is parallel to the $\Vec{d}$-vector. 
Thus, we conclude that the topologically-protected zero-energy states are equivalent to the Andreev bound states in all topological odd-parity gap functions within the quasiclassical treatment. 

We can confirm the above statement by solving Eq.~(\ref{eq:quasi}). 
A boundary condition for Eq.~(\ref{eq:quasi}) is derived from the original BdG equations with the approximated wave function in Eq.~(\ref{eq:appro}). 
In the quasiclassical treatment, the bound states consist of the linear combination of the wave functions with 
$\Vec{p}_{\rm F in}$ and $\Vec{p}_{\rm F out}$. 
According to Ref.~\citen{Hsieh,NagaiPhysicaC}, we adopt the boundary condition for a surface at $z = 0$ given by $\gamma^{3} \Vec{u}(z = 0) = i \Vec{u}(z = 0)$ and $\gamma^{3} \Vec{u}_{\rm c}(z = 0) = i \Vec{u}_{\rm c}(z = 0)$.
The boundary condition for Eq.~(\ref{eq:quasi}) with $p_{{\rm F} x} = p_{{\rm F} y}  = 0$ becomes $\Vec{f}(
z = 0,p_{{\rm F} z} ) = A(p_{{\rm F} z} ) \Vec{f}(
z = 0,-p_{{\rm F} z} )$ and $\Vec{g}(
z = 0,p_{{\rm F} z} ) = A(p_{{\rm F} z} ) \Vec{g}(
z = 0,-p_{{\rm F} z} )$, where $A(p_{{\rm F} z} ) \equiv (-i p_{{\rm F} z} + E_{0} + M_{0})/(-i p_{{\rm F} z} -(E_{0} + M_{0}))$. 
With the straightforward calculation, we can obtain the result that the zero-energy bound states appear when 
$\hat{\Delta}_{\rm eff}(p_{{\rm F} z}) = -\hat{\Delta}_{\rm eff}(-p_{{\rm F} z})$, which is equivalent to that in the conventional quasiclassical theory. 

%
%
The next issue is the quasiparticle bound states in a vortex core. 
A vortex locally breaks a superconducting order parameter so that quasiparticles form discrete energy levels 
inside a vortex.  
These bound states are also called the Andreev bound states. 
In topological superconductors, the zero-energy bound states called the Majorana bound states appear inside a vortex core. 
There is the useful method to investigate the energy levels 
on the basis of the Bohr-Sommerfeld quantum condition around a vortex\cite{Volovik}. 
%
%
%
In this paper, we apply this method for $\Vec{p}$-wave superconductivity to a topological superconductor. 
We already obtained the zero energy Majorana bound states by 
solving the original Dirac-BdG equations (\ref{eq:bdgdirac}) in the pseudo-scalar type gap function as in Ref.~\citen{NagaiMajo}. 
Here, we show that the pseudo-scalar type has zero-energy Majorana fermions inside the vortex core in terms of the quasiclassical treatment. 
The pseudo-scalar gap whose matrix elements are written in the present quasiclassical theory
as
\begin{equation}
\hat{\Delta}_{\rm eff}^{\rm P-scalar} (\Vec{r},\Vec{p}_{\rm F})= \frac{\Delta_{0}(\Vec{r} )}{E_{0}}\left(\begin{array}{cc}
-p_{x}+ i p_{y} & p_{z} \\
p_{z} & p_{x} + i p_{y}
\end{array}\right). \label{eq:pscalar}
\end{equation}
In the case of $p_{z} = 0$, one can decouple the quasiclassical BdG (Andreev) equations (\ref{eq:quasi}) with 
two chiral $p$-wave gap functions $\Delta_{{\rm eff} \pm} \propto p_{x} \pm \imu p_{y}$.  
In such a case, according to Ref.~\citen{Volovik}, the energy spectrum specified by the vortex line along the $z$-direction is given as %
\begin{eqnarray}
E_{n}(p_{z} = 0) = \omega_{0} n, \label{eq:en} \\
\omega_{0} \equiv \frac{\int_{- \infty}^{\infty} ds \frac{\Delta_{0}(|s|)}{|p_{\rm F}||s|} e^{- 2 K(s)}}{\int_{-\infty}^{\infty} ds e^{- 2 K(s)}},
\end{eqnarray}
where $K(s) = \int_{0}^{s} ds' {\rm sign} s' \Delta_{0}(|s'|)/|\Vec{v}_{\rm F}|$.  
Here, $n$ is the integer quantum number related to the angular momentum.
Equation (\ref{eq:en}) clearly indicates the zero-energy states when $n = 0$. 

The correspondence between the odd-parity topological superconductors and the spin-triplet ones 
is quite useful for finding novel phenomena. 
For example, by using the previous insights on the chiral $p$-wave superconductors\cite{Matsumoto},  we can show that the Majorana vortex bound states are spin polarized in the pseudo-scalar type gap function, which has been 
proposed in our previous paper with the use of the Dirac BdG equations\cite{NagaiMajo}. 
In the case of $p_{z} = 0$, there are the equations for $(f_{1},g_{1})$ with the effective order parameter $p_{x} - i p_{y}$ and for $(f_{2},g_{2})$ with the effective order parameter $p_{x} + i p_{y}$ as shown in Eq.~(\ref{eq:pscalar}). 
We note that $f_{1}$ ($f_{2}$) is the spin-up (spin-down) dominant solution as shown in Eq.~(\ref{eq:ui}). 
Thus, we can consider the effective two-$p$-wave model to discuss this topological superconductivity. 
One can show that the bound states are spin-polarized since the bound states in each $p$-wave state have the different spatial distribution around a vortex due to the internal orbital angular momentum of a $p$-wave Cooper pair, as mentioned in Ref.~\citen{NagaiMajo}. 
%
We also show that the other gap function also has the spin-polarized Majorana bound states. 
In the $z$-polar topological superconductor, the effective gap function is written as 
\begin{equation}
\hat{\Delta}_{\rm eff}^{z{\rm-polar}} (\Vec{r},\Vec{p}_{\rm F})= \frac{\Delta_{0}(\Vec{r} )}{E_{0}} \left(\begin{array}{cc}
\imu (p_{x}- \imu p_{y}) & 0 \\
0 & \imu (p_{x} + \imu p_{y})
\end{array}\right).
\end{equation}
In the magnetic field parallel to the $z$-direction, the present quasiclassical equations can be regarded as 
those with two chiral $p$-wave gap functions $\Delta_{\rm eff} \propto p_{x} \mp \imu p_{y}$, Thus, 
the zero energy Majorana bound states are predicted to have down-spin polarization in the vortex core.

In conclusion, 
we showed that the topologically-protected zero-energy states in a topological superconductor are 
equivalent to the Andreev bound states in terms of the quasiclassical treatment. 
We derived the quasiclassical BdG equations in topological superconductors having strong spin orbit coupling from the Dirac-BdG equations. 
The obtained equations are equivalent to the linearized BdG (Andreev) equations with the effective gap functions shown in Table \ref{table:1}. 
With the use of these effective gap functions, one can easily investigate the topological superconductors 
through the correspondence to 
the spin-triplet superconductors. 
Indeed, we confirmed that the various insights about the two-band behavior of the energy spectrum, the zero-energy surface states,  and the spin-polarized vortex core can be described in the present quasiclassical theory. 
One can use 
several well-developed 
techniques of the quasiclassical theory to study the inhomogeneous topological superconductors with the strong spin-orbit coupling. 

We thank N. Hayashi and K. Tanaka for helpful discussions and comments. 
This study has been supported by Grants-in-Aid for Scientific Research from the Ministry of Education, Culture, Sports, Science and Technology of Japan.




\begin{thebibliography}{9}
\bibitem{ref}
We note that the momentum-dependent $\Vec{d}$-vector directions themselves 
are equivalent to those in the bulk given in Refs.~\citen{Hashimoto} and \citen{Yip}.
\end{thebibliography}
\end{document}